\documentclass[twocolumn,floatfix,superscriptaddress,amsmath,showpacs,showkeys,aps,prb]{revtex4}

\usepackage{t1enc}
\usepackage[final]{graphicx}
\usepackage{graphicx}
\usepackage{epsfig}
\usepackage{bm}
\usepackage[normalem]{ulem}
\usepackage{color}

\begin{document}
%\preprint{APS/123-QED}

\title{Disorder-induced induced mechanism for positive exchange bias fields}

\author{Orlando V. Billoni}
\email{billoni@famaf.unc.edu.ar}
\affiliation{Facultad de Matem\'atica, Astronom\'{\i}a y
F\'{\i}sica, Universidad Nacional de C\'ordoba and \\ Instituto de
F\'{\i}sica Enrique Gaviola  (IFEG-CONICET), Ciudad Universitaria,
5000 C\'ordoba, Argentina}

\author{Francisco A. Tamarit}
%\email{tamarit@famaf.unc.edu.ar}
\affiliation{Facultad de Matem\'atica, Astronom\'{\i}a y
F\'{\i}sica, Universidad Nacional de C\'ordoba and \\ Instituto de
F\'{\i}sica Enrique Gaviola  (IFEG-CONICET), Ciudad Universitaria,
5000 C\'ordoba, Argentina}

\author{Sergio A. Cannas}
%\email{cannas@famaf.unc.edu.ar}
\affiliation{Facultad de Matem\'atica, Astronom\'{\i}a y
F\'{\i}sica, Universidad Nacional de C\'ordoba and \\ Instituto de
F\'{\i}sica Enrique Gaviola  (IFEG-CONICET), Ciudad Universitaria,
5000 C\'ordoba, Argentina}

\date{\today}

\begin{abstract}

We propose a mechanism to explain the phenomenon of positive exchange bias on magnetic bilayered systems.
The mechanism is based on the formation of a domain wall at a disordered interface during field cooling (FC)
which induces a symmetry breaking of the antiferromagnet, without relying on any {\it ad hoc} assumption
about the  coupling between the ferromagnetic (FM) and antiferromagnetic (AFM) layers. The domain wall is a result
of the disorder at the interface between FM and AFM, which reduces the effective anisotropy in the region.
We show that the proposed mechanism explains several known experimental facts within a single theoretical framework.
This result is supported by Monte Carlo simulations on a microscopic Heisenberg model, by micromagnetic
calculations at zero temperature and by mean field analysis of an  effective Ising like phenomenological model.

\end{abstract}

\pacs{75.70.-i, 75.60.Jk, 75.70.Cn}
\keywords{Exchange Bias, Magnetic bilayers.}
%75.70.Cn   Magnetic properties of interfaces (multilayers, superlattices, heterostructures)
%75.60.Jk   Magnetization reversal mechanisms
%75.70.-i   Magnetic properties of thin films, surfaces, and interfaces
\maketitle
%\section{Introduction} \label{intro}
The exchange bias phenomenon\cite{Meiklejohn56PR} (EB)
usually appears in heterogeneous magnetic systems in the nanoscale range,
such as thin-film layered systems. EB has captured the  attention of many researchers
due to its applications\cite{Dieny91PRB,Roy12ACMP}, which make the area an active field
of research\cite{Berkowitz99JMMM, Nogues05PR, Giri11JPCM}. For instance, EB 
is currently applied in the design of spin valves \cite{Dieny91PRB,Radu12NC}. 
The phenomenon manifests itself when the system is cooled down in the presence
of a magnetic field, provided the starting temperature is above a certain
threshold and the final temperature is low enough.
An hysteresis loop performed after this procedure shows an horizontal
shift called the bias field, $H_{EB}$.
Usually the bias field is opposite to the cooling field (normal exchange bias, NEB)
but sometimes the displacement is in the same direction
and it is called positive EB\cite{Nogues96PRL} (PEB).
Other important effects can appear, such as a vertical shift in the magnetization \cite{Nogues00PRB}
and the widening  and symmetry loss of the hysteresis loops. EB disappears
if the system is heated above the blocking temperature, $T_B$,
which is below but close to the Neel temperature of the antiferromagnetic (AFM).
Currently, much of the effort is focused on tuning EB and
establishing the mechanisms which control the effect.
The existence of uncompensated domains at the interface has been
shown to be fundamental for the appearance of EB\cite{Takano97PRL}.
Also, at the relevant scales of the problem all the systems have
some unavoidable amount of disorder which seems to play a main role.
In this regard, several routes are  employed in experiments to introduce
and control the disorder effects  \cite{Nogues96PRL, Shi03JAP, Hong06PRL, Cheon07JAP, Cheon07APL, Munbodh11PRB}.
For instance, dilution can  enhance the bias field.\cite{Hong06PRL} 
In addition, the interfacial roughness and the disorder in the anisotropy, are related to
the appearance of PEB\cite{Nogues96PRL,Cheon07JAP}.
In any case, it is well established that a strong cooling field
is necessary for the observation of PEB\cite{Nogues96PRL,Nogues00PRB,Leighton99PRB}.
Among the bilayered systems, one of the most studied is the FM/FeF$_2$,
because the AFM FeF$_2$ has a simple spin structure\cite{Nogues99PRB}.
In particular, PEB was reported for the first time by Nogu\'es et al.
\cite{Nogues96PRL} in this kind of systems.

Most of the theoretical works up to now has assumed that the AFM/FM interface exchange
interaction is antiferromagnetic; this is a key ingredient to explain PEB\cite{Koon97PRL,Schulthess98PRL,Kiwi99EL,Kiwi99APL,Kiwi00SSC,Nowak02PRB,Hu11PSSb}.
In this paper we show that such an {\it ad hoc} assumption (hard to justify physically) is
not necessary to explain PEB, as long as a large enough amount of disorder is
present at the interface. To exhibit the mechanism behind such an effect we first
performed Monte Carlo simulations using a microscopy model for the bilayered system.
We show that PEB is related to the formation of a domain wall at the interface
during field cooling (FC), hence in this case PEB is independent of the sign of the
interface exchange interaction.
%

%Model
We considered a FM film mounted over an AFM film. The films are magnetically coupled
to each other by exchange interactions and the structure of both films is $bcc$, assuming a perfect
match across the FM/AFM interface. The system is ruled by the following Hamiltonian,
\begin{widetext}
\begin{eqnarray}
H &=&  -J_F\!\!\!\!\!\! \sum_{<\vec{r},\vec{r}'> \in \mbox{\footnotesize FM}} \vec{S}_{\vec{r}} \cdot \vec{S}_{\vec{r}'} - K_F \sum _{\vec{r} \in \mbox{\footnotesize FM}} (S^z_{\vec{r}})^2
+ J_{A} \sum _{<\vec{r},\vec{r}'> \in \mbox{\footnotesize AFM}}  \vec{S}_{\vec{r}} \cdot \vec{S}_{\vec{r}'}
- K_A  \sum _{\vec{r} \in \mbox{\footnotesize AFM}} (\vec{S}_{\vec{r}}\cdot \hat{n}_{\vec{r}})^2 \nonumber \\
& &-J_{EB}\!\!\!\!\!\!\!\!\!\!\! \sum _{<\vec{r},\vec{r}'> \in \mbox{\footnotesize FM/AFM}} \vec{S}_{\vec{r}} \cdot \vec{S}_{\vec{r}'}
 - h \sum _{\vec{r}} S^y_{\vec{r}},
\label{eq1}
\end{eqnarray}
\end{widetext}
\noindent where $\vec{S}_{\vec{r}}$ is a classical Heisenberg
spin ($|\vec{S}_{\vec{r}}|=1$) located at the node $\vec{r}$ of the lattice.
$<\vec{r},\vec{r}\,'>$ denotes a sum over nearest-neighbors
pairs of spins. $J_F>0$ is the exchange constant of the FM,  $J_{A}>0$ is
the strength of the AFM exchange, and $J_{EB}>0$ is the exchange coupling between
the FM and the AFM at the interface.
$K_F$ and $K_A$ are FM and AFM anisotropy constants, respectively.
The disorder in the anisotropy is introduced as in the random anisotropy
model \cite{Billoni05PRB}, i.e. $\hat{n}_{\vec{r}}$ is a random direction versor
for AFM spins close to the interface.
Inside the AFM $\hat{n}_{\vec{r}}$ points in the $y$ direction.
$h$ is an external homogeneous magnetic field oriented along
the $y$ direction. We assumed $K_F<0$ --planar anisotropy-- to ensure the FM spins
remain in the film plane, mimicking the dipolar shape anisotropy\cite{Nowak02JMMM}.
Monte Carlo  simulations were performed using  Metropolis algorithm.
$L_x = L_y = L$ are the lateral dimensions of the films,
and $L_{za}$ and $L_{zf}$ are the number of atomic layer of the FM and AFM films,
respectively.
We set $L_x=L_y=20,40$, $L_{za}=24$ and $L_{zf}=12$.
Periodic boundary conditions were imposed in the plane of the film while
open boundary conditions were used in the perpendicular direction.
For each point in the magnetization curve, we took
$10^4$ Monte Carlo steps per site (MCS) to thermalize the system and the
same number of MCS to calculate temporal averages.
The AFM was modeled using FeF$_2$ fluoride parameters
\footnote{We fixed the  parameters to be: $J_F=9.56J$, $J_A=-J$,
and $K_F=-0.5J$, where $J$ is an arbitrary parameter that sets the
energy units.  $J_{EB}=0.5J,J$ and $K_A=1.77J$.}
setting AFM interface spin configuration  uncompensated \cite{Billoni06PB,Lederman04PRB,Billoni11JPCM},
corresponding to the (100) FeF$_2$ crystalline orientation.
%

%
%\subsection{Results}
%\label{res}
%

\begin{figure}%[ht]
\includegraphics*[width=8cm,angle=0]{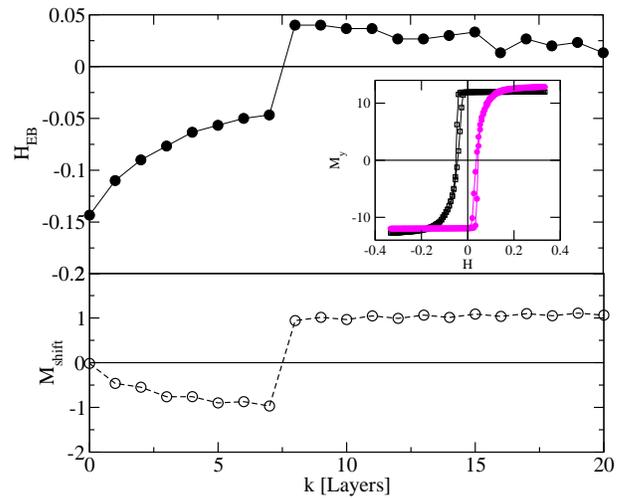}
\caption{\label{fig1} (Color online) $H_{EB}$ and $M_{shift}$ vs. the number
of disordered planes at low temperature ($T/T_N \sim 0.1$ and $H_{CF}=0.32$)
top and bottom panel, respectively.
Inset: hysteresis loops before and after the change of sign of $H_{EB}$.
Note that $M_{shift}$ stabilizes to a value close to 1. This is expected according to
Eq. (\ref{mag}).
}
\end{figure}

\begin{figure}%[ht]
\includegraphics*[width=8cm,angle=0]{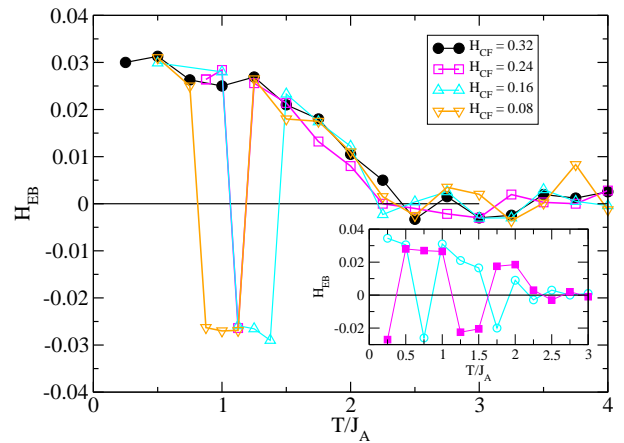}
\caption{\label{fig2} (Color online) Bias Field as function of the temperature.
The AFM film includes 12 layers with disorder. Different symbols correspond
to different cooling fields $L=40$. Inset: two samples in a reduced system size
$L=20$ $H_{CF}=0.08$
}
\end{figure}

%
In Fig.\ref{fig1}, we show the field $H_{EB}$ versus the number of disordered 
AFM layers $k$ in the interface region.
The temperature of the system is well below the N\`eel temperature ($T/T_N=0.1$).
As a general rule the bias field decreases in module as the number of planes
with disorder increases, as expected according to previous results
(see e.g. Ref.\onlinecite{Kim01APL}).
The bias field is normal up to $k=7$, and for larger values it becomes positive.
Notice that the  absolute value of $H_{EB}$ varies continuously at the transition
from normal to PEB, as observed experimentally\cite{Nogues96PRL}. We also observe
a vertical shift in the hysteresis loops correlated with the sign of
the bias field (see Fig.\ref{fig1}), as observed in fluoride iron
compounds \cite{Nogues00PRB}.
In Fig.\ref{fig2} we plot $H_{EB}$ versus temperature for a system with
a fixed number of AFM disordered layers (k=12) and for different cooling fields.
If the cooling field is strong ($H_{CF} > 0.24J_A$), $H_{EB}$ is positive in the
whole range of temperatures whereas, as $H_{CF}$ decreases,
the sign of $H_{EB}$ changes twice at intermediate temperatures.
A change of sign of $H_{EB}$ as a function of the temperature has been observed in diluted
AFM \cite{Shi03JAP} and in random anisotropy AFM \cite{Cheon07JAP}. As we will discuss later,
while NEB is expected at low temperatures, the presence of PEB at very low temperatures
appears to be a spurious finite size effect, as suggested by the strong enhancement of fluctuations
in the sign of $H_{EB}$ as the lateral size of the system is reduced (see inset of Fig.\ref{fig2}).
%Note that the range where $H_{EB}$ becomes normal seems to increase as $H_{CF}$ decreases.
%

An inspection of the local magnetization at each layer shows that,
in the case of PEB, an antiferromagnetic domain wall (DW) forms during
FC in the disordered region.
In this way the system reduces the exchange energy cost due to frustration
while it stores energy at the interface through the Zeeman coupling of the AFM
spins. This energy is restored during the field reversal producing a positive bias
in the hysteresis loop. On the contrary, in the case of NEB no DW is observed
for positive field. In other words, the whole AFM slab (both the ordered and the
disordered regions), exhibits a single Neel state without frustration. In this
case, a DW forms for {\em negative} fields, giving rise to a negative bias
of the hysteresis loop. It is noting to note that in our simulations, 
the DW formed in the disordered region is responsible for the shift in the 
magnetization.

\begin{figure}%[ht]
\includegraphics*[width=8cm,angle=0]{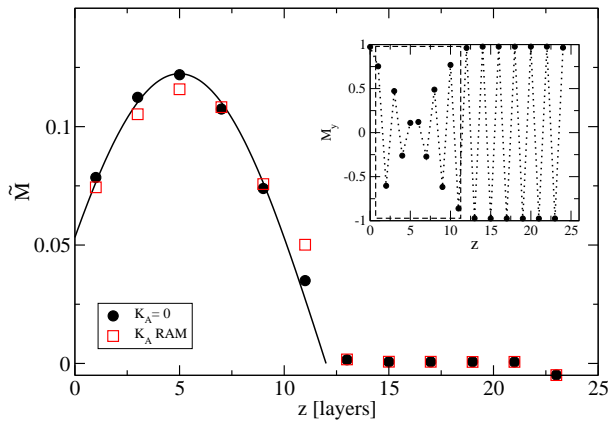}
\caption{\label{fig3} (Color online) Net magnetization of the AFM $\tilde{M}(z)$
obtained through Monte Carlo simulations, RAM refers to random anisotropy.
The solid line corresponds to Eq.(\ref{mag}) with $\theta_0=0.45$ and $l_w=11$.
Inset: AFM magnetization profile, the dashed box indicates the disordered
region.
}
\end{figure}

Let's analyze the conditions for the formation of a domain wall under
an applied field in the  disordered region of the AFM.
The energy per unit of area of an AFM disordered region of length $l_w$ under
the applied field is\footnote{See Supplemental Material at 
http://link.aps.org/supplemental/10.1103/PhysRevB.88.020405 for more details about these  calculations.}.

\begin{equation}
E=\int_0^{l_w} \left[ \frac{J_l}{2}\left(\frac{d\theta}{dz}\right)^2 -
\frac{H_{CF}}{2}\sin(\theta)\left(\frac{d\theta}{dz}\right)\right] dz.
\end{equation}

Assuming that randomness averages the effect of the anisotropy, we neglected it
considering only the exchange interaction between layers ($J_l$) and the coupling with the field.
We will test this approximation later. Note the field interacts through the gradient of the angle
$\theta$ since this region is antiferromagnetically ordered, i.e. the magnetization
in the direction of the field is $m(z) = \frac{1}{2}\sin(\theta)\left(\frac{d\theta}{dz}\right)$.
Minimizing this energy we obtain $\theta  =  \frac{\pi-\theta_0}{l_w} z + \theta_0$,
where  $\theta_0$ is a free parameter ($0<\theta_0<\pi$). The total energy is,
\begin{equation}
E= \frac{J_l}{2}\left(\frac{\pi-\theta_0}{l_w}\right)^2 l_w - \frac{H_{CF}}{2}\left(1 + \cos(\theta_0)\right),
\end{equation}

\noindent and a domain wall forms if $E<0$ implying $H_{CF} > H^*$, where
\begin{equation}
\label{hcf}
H^* = J_l \frac{(\pi-\theta_0)^2}{[1 + \cos(\theta_0)]l_w}.
\end{equation}

\noindent The magnetization profile is:

\begin{equation}
\label{mag}
m(z)= \frac{1}{2}\frac{\pi-\theta_0}{l_w}\sin\left(\frac{\pi-\theta_0}{l_w}z + \theta_0\right).
\end{equation}

We checked Eq.(\ref{mag}) simulating (Eq.(\ref{eq1}))  a system containing a region
without anisotropy.
In Fig. \ref{fig3} we plot the profile of the net magnetization  $\tilde{M} (z)=[m_y(z)+m_y(z+1)]/2$
pointing in the direction of the applied field. The continuous
line is a fit of Eq.(\ref{mag}) with $\theta_0=0.45$, showing a good agreement
with the Monte Carlo simulation.
Moreover, when randomness is considered ($K_A \ne 0$),
the agreement is still good, verifying our previous assumption.

According to Eq.(\ref{hcf}) there is a threshold
for the appearance of a domain wall and therefore PEB. Using the parameters of the simulations (Fig. \ref{fig3})
and $J_l=4J_A$ we obtain  $H^* = 1.38J_A$ ($\sim 5.5$ T)\footnote{The molecular field of FeF$_2$ is $8J_A\sim 32$ T.}.
However in simulations PEB is observed at cooling
fields as low as $0.08J_A (\sim 0.32$ T). To explain this discrepancy one has to assume
the domain wall forms at higher temperatures where $J_l \ll 4J_A$. This is plausible since
PEB is observed in our simulations even at temperatures close to $T_N$ \footnote{The blocking temperatures in Fig. \ref{fig2} 
are about 0.9 the N\`eel temperature.}.

To analyze thermal effects in the DW formation mechanism we consider a phenomenological model.
Assuming that the spins at each AFM layer behave coherently, we associate
an Ising spin $\sigma_i=\pm 1$ ($i=1,\ldots,l+L$), corresponding to the magnetization per unit  area component in the
direction of the applied field for the layer $i$. The magnetization per unit area of the FM slab is represented
by $S= t_F\, \sigma_0$, where $t_F$ is the thickness of the FM slab and $\sigma_0=\pm 1$. The disordered interface
is represented by the first $l$ layers ($i=1,\ldots,l$). We assume that the anisotropy at the ordered
region  $i=l+1,\ldots,l+L$ ($L \gg l$) is very strong, so that the Zeeman contribution of that region can be
neglected. On the other hand, we assume the anisotropy at the disordered region can be neglected, compared with
the corresponding Zeeman term. Then, the Hamiltonian for the effective model is given by

\begin{equation}\label{HIsing}
{\cal H} = - Jt_F  \sigma_0\sigma_1  +   J_{AF}\, \sum_{i=1}^{l+L-1} \sigma_i\sigma_{i+1}- H t_F \sigma_0  -H  \sum_{i=1}^l\sigma_{i}
\end{equation}

\noindent with $J>0$ and $J_{AF}>0$. At zero temperature, a simple analysis of Eq.(\ref{HIsing}) shows the existence
of a threshold $H^*=J_{AF}$  such that a DW forms only when $H>H^*$, consistently with the previous micromagnetic
calculation.

\begin{figure}%[ht]
\includegraphics*[width=6cm,angle=-90]{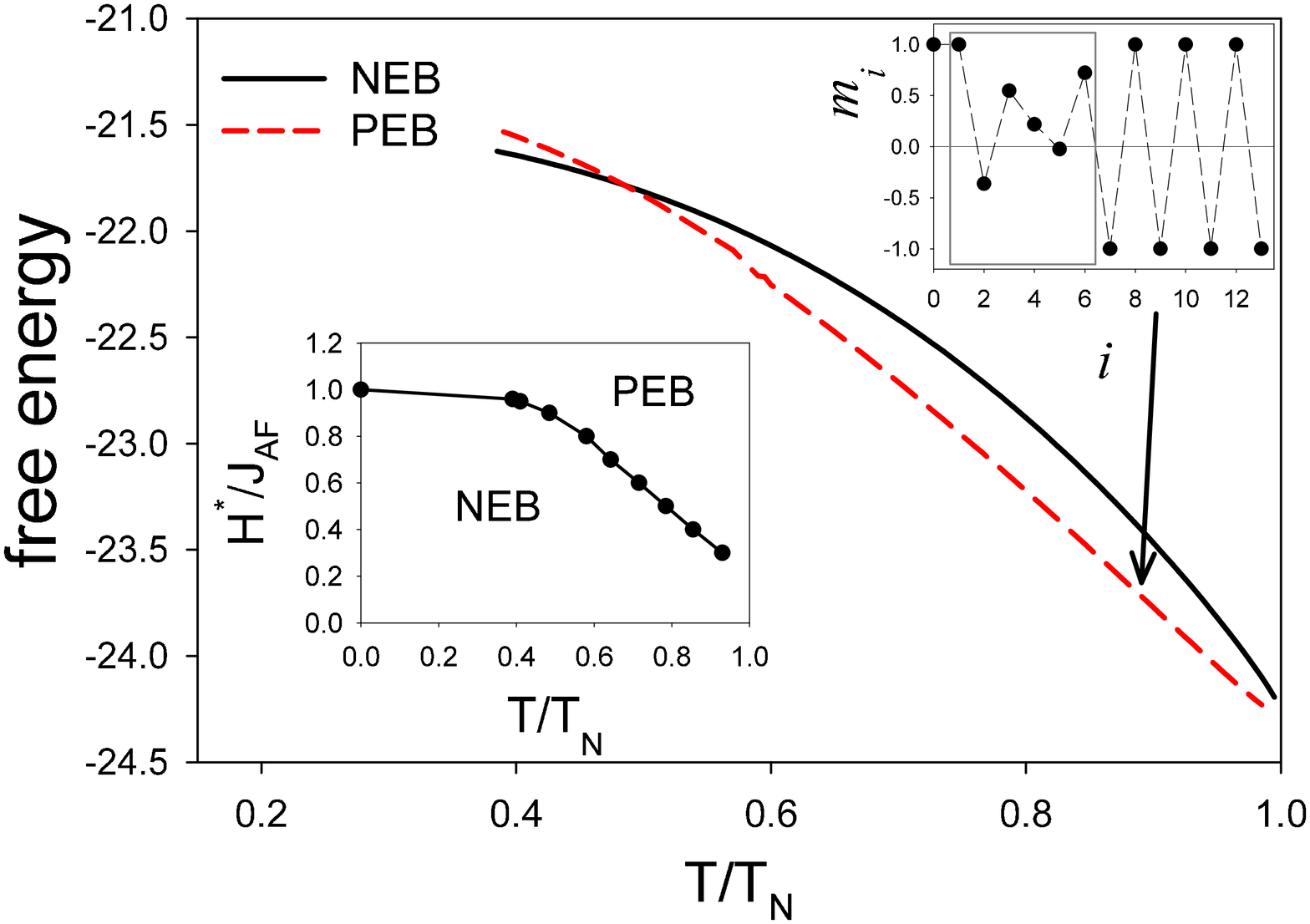}
\caption{\label{fig4} (Color online) Mean field free energy for Hamiltonian (\ref{HIsing})
as a function of the temperature for $H=0.9$ and $l=6$ corresponding to PEB (red dashed line)
and NEB (continuous black line) AFM configurations. Upper inset: Local magnetization at the PEB  
for $T=0.85T_N$. Lower inset: minimum cooling field for having PEB as a function of the temperature.
}
\end{figure}

At finite temperature, a variational mean-field free energy can be easily 
derived\cite{Chaikin95}$^,$ 
\footnote{See Supplemental Material at http://link.aps.org/supplemental/10.1103/PhysRevB.88.020405 
for more details about the derivation of the free energy and the calculations described 
in what follows.}
in terms of the local average magnetizations $m_i = \langle \sigma_i \rangle$ ($i=0,\ldots\infty$).
During FC, the ordered AFM slab ($i>l$)  takes a configuration that minimizes the whole free energy.
Such a configuration  remains fixed when the field is retired at low temperatures, while the disordered AFM
region and the FM slab (i.e., those spins which interact with the field) are capable of  accommodating
a new minimum free energy configuration. The analysis can be further simplified by assuming that the
ordered AFM conforms to a Neel state with the local sublattice magnetization given by the  Curie
equation $m_{AF} = \tanh[2 \beta J_{AF} m_{AF}]$ ($\beta=1/k_B T$). We then have two different possibilities
(let us assume for simplicity that $l$ is even): {\it i}) $m_{l+1}= m_{AF}$ (we are considering the positive
root of the previous Curie equation and assuming $T< T_N= 2J_{AF}/k_B$). In this case, there is no DW and
$m_F\equiv t_F \langle m_0 \rangle >0$ when $H=0$, thus corresponding to  NEB.  {\it ii}) $m_{l+1}= -m_{AF}$,
there is a DW and $m_F <0$ when $H=0$, thus corresponding to  PEB. We numerically obtained the minimum
free energy solution for both possibilities and compared them for different values of $H$ and $T$.
In Fig.\ref{fig4} we show a typical example of both free energies when\footnote{We used in the calculations
$J_{AF}=1$, $J=J_{AF}/2$ and $t_F=5$.} $H< J_{AF}$. We see that the minimum free energy solution changes from
NEB to PEB as the temperature increases.  Conversely, for each temperature we have a minimum field $H^*(T)$
(see lower inset of Fig.\ref{fig4}) such that PEB becomes the minimum free energy state when $H>H^*(T)$,
even for temperatures close to $T_N$. The upper inset of Fig.\ref{fig4} shows the DW in the PEB case
(compare with the inset of Fig.\ref{fig3}). A change in the sign of $H_{EB}$ as a function of the
temperature has been observed in disordered fluorides \cite{Shi03JAP,Cheon07APL,Cheon07JAP,Munbodh11PRB}
following the same trend we observed in the MF calculations.
In particular, in the Fe$_x$Ni$_{1-x}$F$_2$/Co  bilayer\cite{Munbodh11PRB} a domain wall at the interface has
been reported, where Fe$_x$Ni$_{1-x}$F$_2$ is a random anisotropy antiferromagnet. Since the critical field $H^*$ depends
on the amount of disordered layers, inhomogeneities at the interface can give rise to a distribution of $H^*$
as is observed in FeF$_2$/FM \cite{Petracic05APL,Roshchin05EPL}.
Finally, the effect of the disorder of the anisotropy at the interface is similar to that considered 
in the spin glass model of exchange bias\cite{Radu08}, since disorder reduces
the anisotropy at the interface,  but in our case the coupling of this region with the applied 
field turns out to be important to produce PEB.  

Summarizing, the reduction in the anisotropy for a large enough amount of interfacial disorder can induce the
formation of a domain wall in the cooling field process inducing a symmetry breaking in the antiferromagnet. The energy
stored in this domain wall is released during the field reversion, resulting in PEB. In this way, the PEB 
phenomenon can be explained as an exclusive result of interfacial disorder, without relying on ad hoc assumptions 
about the sign of the coupling between FM and AFM.
This work was partially supported by grants from CONICET, and SeCyT
Universidad Nacional de C\'ordoba (Argentina).

%\bibliographystyle{elsart-num}
%\bibliographystyle{apsrev}
%\bibliography{ebias,old_ebias}
%\bibliography{old_ebias}

\begin{thebibliography}{34}
\expandafter\ifx\csname natexlab\endcsname\relax\def\natexlab#1{#1}\fi
\expandafter\ifx\csname bibnamefont\endcsname\relax
  \def\bibnamefont#1{#1}\fi
\expandafter\ifx\csname bibfnamefont\endcsname\relax
  \def\bibfnamefont#1{#1}\fi
\expandafter\ifx\csname citenamefont\endcsname\relax
  \def\citenamefont#1{#1}\fi
\expandafter\ifx\csname url\endcsname\relax
  \def\url#1{\texttt{#1}}\fi
\expandafter\ifx\csname urlprefix\endcsname\relax\def\urlprefix{URL }\fi
\providecommand{\bibinfo}[2]{#2}
\providecommand{\eprint}[2][]{\url{#2}}

\bibitem[{\citenamefont{Meiklejohn and Bean}(1956)}]{Meiklejohn56PR}
\bibinfo{author}{\bibfnamefont{W.~H.} \bibnamefont{Meiklejohn}}
  \bibnamefont{and} \bibinfo{author}{\bibfnamefont{C.~P.} \bibnamefont{Bean}},
  \bibinfo{journal}{Phys. Rev.} \textbf{\bibinfo{volume}{102}},
  \bibinfo{pages}{1413} (\bibinfo{year}{1956}).

\bibitem[{\citenamefont{Dieny et~al.}(1991)\citenamefont{Dieny, Speriosu,
  Parkin, Gurney, Wilhoit, and Mauri}}]{Dieny91PRB}
\bibinfo{author}{\bibfnamefont{B.}~\bibnamefont{Dieny}},
  \bibinfo{author}{\bibfnamefont{V.~S.} \bibnamefont{Speriosu}},
  \bibinfo{author}{\bibfnamefont{S.~S.~S.} \bibnamefont{Parkin}},
  \bibinfo{author}{\bibfnamefont{B.~A.} \bibnamefont{Gurney}},
  \bibinfo{author}{\bibfnamefont{D.~R.} \bibnamefont{Wilhoit}},
  \bibnamefont{and} \bibinfo{author}{\bibfnamefont{D.}~\bibnamefont{Mauri}},
  \bibinfo{journal}{Phys. Rev. B} \textbf{\bibinfo{volume}{43}},
  \bibinfo{pages}{1297} (\bibinfo{year}{1991}).

\bibitem[{\citenamefont{Roy et~al.}(2012)\citenamefont{Roy, Gupta, and
  Garg}}]{Roy12ACMP}
\bibinfo{author}{\bibfnamefont{A.}~\bibnamefont{Roy}},
  \bibinfo{author}{\bibfnamefont{R.}~\bibnamefont{Gupta}}, \bibnamefont{and}
  \bibinfo{author}{\bibfnamefont{A.}~\bibnamefont{Garg}},
  \bibinfo{journal}{Adv. Condens. Matter Phys.} \textbf{\bibinfo{volume}{2012}},
  \bibinfo{pages}{Article ID 926290} (\bibinfo{year}{2012}).

\bibitem[{\citenamefont{Berkowitz and Kentaro}(1999)}]{Berkowitz99JMMM}
\bibinfo{author}{\bibfnamefont{A.~E.} \bibnamefont{Berkowitz}}
  \bibnamefont{and} \bibinfo{author}{\bibfnamefont{T.}~\bibnamefont{Kentaro}},
  \bibinfo{journal}{J. Magn. Magn. Mater.} \textbf{\bibinfo{volume}{200}},
  \bibinfo{pages}{552} (\bibinfo{year}{1999}).

\bibitem[{\citenamefont{Nogu\'es et~al.}(2005)\citenamefont{Nogu\'es, Sort,
  Langlais, Skumryev, Suri\~nach, Mu\~noz, and Bar\'o}}]{Nogues05PR}
\bibinfo{author}{\bibfnamefont{J.}~\bibnamefont{Nogu\'es}},
  \bibinfo{author}{\bibfnamefont{J.}~\bibnamefont{Sort}},
  \bibinfo{author}{\bibfnamefont{V.}~\bibnamefont{Langlais}},
  \bibinfo{author}{\bibfnamefont{V.}~\bibnamefont{Skumryev}},
  \bibinfo{author}{\bibfnamefont{S.}~\bibnamefont{Suri\~nach}},
  \bibinfo{author}{\bibfnamefont{J.~S.} \bibnamefont{Mu\~noz}},
  \bibnamefont{and} \bibinfo{author}{\bibfnamefont{M.~D.}
  \bibnamefont{Bar\'o}}, \bibinfo{journal}{Phys. Rep.}
  \textbf{\bibinfo{volume}{422}}, \bibinfo{pages}{65} (\bibinfo{year}{2005}).

\bibitem[{\citenamefont{Giri et~al.}(2011)\citenamefont{Giri, Patra, and
  Majumdar}}]{Giri11JPCM}
\bibinfo{author}{\bibfnamefont{S.}~\bibnamefont{Giri}},
  \bibinfo{author}{\bibfnamefont{M.}~\bibnamefont{Patra}}, \bibnamefont{and}
  \bibinfo{author}{\bibfnamefont{S.}~\bibnamefont{Majumdar}},
  \bibinfo{journal}{J. Phys.: Condens. Matter} \textbf{\bibinfo{volume}{23}},
  \bibinfo{pages}{073201} (\bibinfo{year}{2011}).

\bibitem[{\citenamefont{Radu et~al.}(2012)\citenamefont{Radu, Abrudan, Radu,
  Schmitz, and Zabel}}]{Radu12NC}
\bibinfo{author}{\bibfnamefont{F.}~\bibnamefont{Radu}},
  \bibinfo{author}{\bibfnamefont{R.}~\bibnamefont{Abrudan}},
  \bibinfo{author}{\bibfnamefont{I.}~\bibnamefont{Radu}},
  \bibinfo{author}{\bibfnamefont{D.}~\bibnamefont{Schmitz}}, \bibnamefont{and}
  \bibinfo{author}{\bibfnamefont{H.}~\bibnamefont{Zabel}},
  \bibinfo{journal}{Nat Commun} \textbf{\bibinfo{volume}{3}},
  \bibinfo{pages}{715} (\bibinfo{year}{2012}).

\bibitem[{\citenamefont{Nogu\'es et~al.}(1996)\citenamefont{Nogu\'es, Lederman,
  Moran, and Schuller}}]{Nogues96PRL}
\bibinfo{author}{\bibfnamefont{J.}~\bibnamefont{Nogu\'es}},
  \bibinfo{author}{\bibfnamefont{D.}~\bibnamefont{Lederman}},
  \bibinfo{author}{\bibfnamefont{T.~J.} \bibnamefont{Moran}}, \bibnamefont{and}
  \bibinfo{author}{\bibfnamefont{I.~K.} \bibnamefont{Schuller}},
  \bibinfo{journal}{Phys. Rev. Lett.} \textbf{\bibinfo{volume}{76}},
  \bibinfo{pages}{4624} (\bibinfo{year}{1996}).

\bibitem[{\citenamefont{Nogu\'es et~al.}(2000)\citenamefont{Nogu\'es, Leighton,
  and Schuller}}]{Nogues00PRB}
\bibinfo{author}{\bibfnamefont{J.}~\bibnamefont{Nogu\'es}},
  \bibinfo{author}{\bibfnamefont{C.}~\bibnamefont{Leighton}}, \bibnamefont{and}
  \bibinfo{author}{\bibfnamefont{I.~K.} \bibnamefont{Schuller}},
  \bibinfo{journal}{Phys. Rev. B} \textbf{\bibinfo{volume}{61}},
  \bibinfo{pages}{1315} (\bibinfo{year}{2000}).

\bibitem[{\citenamefont{Takano et~al.}(1997)\citenamefont{Takano, Kodama,
  Berkowitz, Cao, and Thomas}}]{Takano97PRL}
\bibinfo{author}{\bibfnamefont{K.}~\bibnamefont{Takano}},
  \bibinfo{author}{\bibfnamefont{R.~H.} \bibnamefont{Kodama}},
  \bibinfo{author}{\bibfnamefont{A.~E.} \bibnamefont{Berkowitz}},
  \bibinfo{author}{\bibfnamefont{W.}~\bibnamefont{Cao}}, \bibnamefont{and}
  \bibinfo{author}{\bibfnamefont{G.}~\bibnamefont{Thomas}},
  \bibinfo{journal}{Phys. Rev. Lett.} \textbf{\bibinfo{volume}{79}},
  \bibinfo{pages}{1130} (\bibinfo{year}{1997}).

\bibitem[{\citenamefont{Shi et~al.}(2003)\citenamefont{Shi, D.~Lederman,
  Dilley, Black, Diedrichs, Jensen, and Simmonds}}]{Shi03JAP}
\bibinfo{author}{\bibfnamefont{H.}~\bibnamefont{Shi}},
  \bibinfo{author}{\bibfnamefont{D.}~\bibnamefont{D.~Lederman}},
  \bibinfo{author}{\bibfnamefont{N.~R.} \bibnamefont{Dilley}},
  \bibinfo{author}{\bibfnamefont{R.~C.} \bibnamefont{Black}},
  \bibinfo{author}{\bibfnamefont{J.}~\bibnamefont{Diedrichs}},
  \bibinfo{author}{\bibfnamefont{K.}~\bibnamefont{Jensen}}, \bibnamefont{and}
  \bibinfo{author}{\bibfnamefont{M.~B.} \bibnamefont{Simmonds}},
  \bibinfo{journal}{J. Appl. Phys.} \textbf{\bibinfo{volume}{93}},
  \bibinfo{pages}{8600} (\bibinfo{year}{2003}).

\bibitem[{\citenamefont{Hong et~al.}(2006)\citenamefont{Hong, Leo, Smith, and
  Berkowitz}}]{Hong06PRL}
\bibinfo{author}{\bibfnamefont{J.-I.} \bibnamefont{Hong}},
  \bibinfo{author}{\bibfnamefont{T.}~\bibnamefont{Leo}},
  \bibinfo{author}{\bibfnamefont{D.~J.} \bibnamefont{Smith}}, \bibnamefont{and}
  \bibinfo{author}{\bibfnamefont{A.~E.} \bibnamefont{Berkowitz}},
  \bibinfo{journal}{Phys. Rev. Lett.} \textbf{\bibinfo{volume}{96}},
  \bibinfo{pages}{117204} (\bibinfo{year}{2006}).

\bibitem[{\citenamefont{Cheon et~al.}(2007{\natexlab{a}})\citenamefont{Cheon,
  Liu, and Lederman}}]{Cheon07JAP}
\bibinfo{author}{\bibfnamefont{M.}~\bibnamefont{Cheon}},
  \bibinfo{author}{\bibfnamefont{Z.}~\bibnamefont{Liu}}, \bibnamefont{and}
  \bibinfo{author}{\bibfnamefont{D.}~\bibnamefont{Lederman}},
  \bibinfo{journal}{J. Appl. Phys.} \textbf{\bibinfo{volume}{101}},
  \bibinfo{eid}{09E503} (\bibinfo{year}{2007}{\natexlab{a}}).

\bibitem[{\citenamefont{Cheon et~al.}(2007{\natexlab{b}})\citenamefont{Cheon,
  Liu, and Lederman}}]{Cheon07APL}
\bibinfo{author}{\bibfnamefont{M.}~\bibnamefont{Cheon}},
  \bibinfo{author}{\bibfnamefont{Z.}~\bibnamefont{Liu}}, \bibnamefont{and}
  \bibinfo{author}{\bibfnamefont{D.}~\bibnamefont{Lederman}},
  \bibinfo{journal}{Appl. Phys. Lett.} \textbf{\bibinfo{volume}{90}},
  \bibinfo{eid}{012511} (\bibinfo{year}{2007}{\natexlab{b}}).

\bibitem[{\citenamefont{Munbodh et~al.}(2011)\citenamefont{Munbodh, Cheon,
  Lederman, Fitzsimmons, and Dilley}}]{Munbodh11PRB}
\bibinfo{author}{\bibfnamefont{K.}~\bibnamefont{Munbodh}},
  \bibinfo{author}{\bibfnamefont{M.}~\bibnamefont{Cheon}},
  \bibinfo{author}{\bibfnamefont{D.}~\bibnamefont{Lederman}},
  \bibinfo{author}{\bibfnamefont{M.~R.} \bibnamefont{Fitzsimmons}},
  \bibnamefont{and} \bibinfo{author}{\bibfnamefont{N.~R.}
  \bibnamefont{Dilley}}, \bibinfo{journal}{Phys. Rev. B}
  \textbf{\bibinfo{volume}{84}}, \bibinfo{pages}{214434}
  (\bibinfo{year}{2011}).

\bibitem[{\citenamefont{Leighton et~al.}(1999)\citenamefont{Leighton, Nogu\'es,
  Suhl, and Schuller}}]{Leighton99PRB}
\bibinfo{author}{\bibfnamefont{C.}~\bibnamefont{Leighton}},
  \bibinfo{author}{\bibfnamefont{J.}~\bibnamefont{Nogu\'es}},
  \bibinfo{author}{\bibfnamefont{H.}~\bibnamefont{Suhl}}, \bibnamefont{and}
  \bibinfo{author}{\bibfnamefont{I.~K.} \bibnamefont{Schuller}},
  \bibinfo{journal}{Phys. Rev. B} \textbf{\bibinfo{volume}{60}},
  \bibinfo{pages}{12837} (\bibinfo{year}{1999}).

\bibitem[{\citenamefont{Nogu\'es et~al.}(1999)\citenamefont{Nogu\'es, m~T.~J.,
  Lederman, Schuller, and Rao}}]{Nogues99PRB}
\bibinfo{author}{\bibfnamefont{J.}~\bibnamefont{Nogu\'es}},
  \bibinfo{author}{\bibfnamefont{T.~J.}~\bibnamefont{Moran}},
  \bibinfo{author}{\bibfnamefont{D.}~\bibnamefont{Lederman}},
  \bibinfo{author}{\bibfnamefont{I.~K.} \bibnamefont{Schuller}},
  \bibnamefont{and} \bibinfo{author}{\bibfnamefont{K.~V.} \bibnamefont{Rao}},
  \bibinfo{journal}{Phys. Rev. B} \textbf{\bibinfo{volume}{59}},
  \bibinfo{pages}{6984} (\bibinfo{year}{1999}).

\bibitem[{\citenamefont{Koon}(1997)}]{Koon97PRL}
\bibinfo{author}{\bibfnamefont{N.~C.} \bibnamefont{Koon}},
  \bibinfo{journal}{Phys. Rev. Lett.} \textbf{\bibinfo{volume}{78}},
  \bibinfo{pages}{4865} (\bibinfo{year}{1997}).

\bibitem[{\citenamefont{Schulthess and Butler}(1998)}]{Schulthess98PRL}
\bibinfo{author}{\bibfnamefont{T.~C.} \bibnamefont{Schulthess}}
  \bibnamefont{and} \bibinfo{author}{\bibfnamefont{W.~H.}
  \bibnamefont{Butler}}, \bibinfo{journal}{Phys. Rev. Lett.}
  \textbf{\bibinfo{volume}{81}}, \bibinfo{pages}{4516} (\bibinfo{year}{1998}).

\bibitem[{\citenamefont{Kiwi et~al.}(1999{\natexlab{a}})\citenamefont{Kiwi,
  Mej\'{\i}a-L\'opez, Portugal, and Ram\'{\i}rez}}]{Kiwi99EL}
\bibinfo{author}{\bibfnamefont{M.}~\bibnamefont{Kiwi}},
  \bibinfo{author}{\bibfnamefont{J.}~\bibnamefont{Mej\'{\i}a-L\'opez}},
  \bibinfo{author}{\bibfnamefont{R.~D.} \bibnamefont{Portugal}},
  \bibnamefont{and}
  \bibinfo{author}{\bibfnamefont{R.}~\bibnamefont{Ram\'{\i}rez}},
  \bibinfo{journal}{Europhys. Lett.} \textbf{\bibinfo{volume}{48}},
  \bibinfo{pages}{573} (\bibinfo{year}{1999}{\natexlab{a}}).

\bibitem[{\citenamefont{Kiwi et~al.}(1999{\natexlab{b}})\citenamefont{Kiwi,
  Mej\'{\i}a-L\'opez, Portugal, and Ram\'{\i}rez}}]{Kiwi99APL}
\bibinfo{author}{\bibfnamefont{M.}~\bibnamefont{Kiwi}},
  \bibinfo{author}{\bibfnamefont{J.}~\bibnamefont{Mej\'{\i}a-L\'opez}},
  \bibinfo{author}{\bibfnamefont{R.~D.} \bibnamefont{Portugal}},
  \bibnamefont{and}
  \bibinfo{author}{\bibfnamefont{R.}~\bibnamefont{Ram\'{\i}rez}},
  \bibinfo{journal}{Appl. Phys. Lett.} \textbf{\bibinfo{volume}{75}},
  \bibinfo{pages}{3395} (\bibinfo{year}{1999}{\natexlab{b}}).

\bibitem[{\citenamefont{Kiwi et~al.}(2000)\citenamefont{Kiwi,
  Mej\'{\i}a-L\'opez, Portugal, and Ram\'{\i}rez}}]{Kiwi00SSC}
\bibinfo{author}{\bibfnamefont{M.}~\bibnamefont{Kiwi}},
  \bibinfo{author}{\bibfnamefont{J.}~\bibnamefont{Mej\'{\i}a-L\'opez}},
  \bibinfo{author}{\bibfnamefont{R.~D.} \bibnamefont{Portugal}},
  \bibnamefont{and}
  \bibinfo{author}{\bibfnamefont{R.}~\bibnamefont{Ram\'{\i}rez}},
  \bibinfo{journal}{Solid State Commun.} \textbf{\bibinfo{volume}{116}},
  \bibinfo{pages}{315} (\bibinfo{year}{2000}).

\bibitem[{\citenamefont{Nowak et~al.}(2002{\natexlab{a}})\citenamefont{Nowak,
  Usadel, Keller, Milt\'enyi, Beschoten, and G\"untherodt}}]{Nowak02PRB}
\bibinfo{author}{\bibfnamefont{U.}~\bibnamefont{Nowak}},
  \bibinfo{author}{\bibfnamefont{K.~D.} \bibnamefont{Usadel}},
  \bibinfo{author}{\bibfnamefont{J.}~\bibnamefont{Keller}},
  \bibinfo{author}{\bibfnamefont{P.}~\bibnamefont{Milt\'enyi}},
  \bibinfo{author}{\bibfnamefont{B.}~\bibnamefont{Beschoten}},
  \bibnamefont{and}
  \bibinfo{author}{\bibfnamefont{G.}~\bibnamefont{G\"untherodt}},
  \bibinfo{journal}{Phys. Rev. B} \textbf{\bibinfo{volume}{66}},
  \bibinfo{pages}{014430} (\bibinfo{year}{2002}{\natexlab{a}}).

\bibitem[{\citenamefont{Hu and Du}(2011)}]{Hu11PSSb}
\bibinfo{author}{\bibfnamefont{Y.}~\bibnamefont{Hu}} \bibnamefont{and}
  \bibinfo{author}{\bibfnamefont{A.}~\bibnamefont{Du}}, \bibinfo{journal}{phys.
  status solidi (b)} \textbf{\bibinfo{volume}{248}}, \bibinfo{pages}{2932}
  (\bibinfo{year}{2011}).

\bibitem[{\citenamefont{Billoni et~al.}(2005)\citenamefont{Billoni, Cannas, and
  Tamarit}}]{Billoni05PRB}
\bibinfo{author}{\bibfnamefont{O.~V.} \bibnamefont{Billoni}},
  \bibinfo{author}{\bibfnamefont{S.~A.} \bibnamefont{Cannas}},
  \bibnamefont{and} \bibinfo{author}{\bibfnamefont{F.~A.}
  \bibnamefont{Tamarit}}, \bibinfo{journal}{Phys. Rev. B}
  \textbf{\bibinfo{volume}{72}}, \bibinfo{pages}{104407}
  (\bibinfo{year}{2005}).

\bibitem[{\citenamefont{Nowak et~al.}(2002{\natexlab{b}})\citenamefont{Nowak,
  Misra, and Usadel}}]{Nowak02JMMM}
\bibinfo{author}{\bibfnamefont{U.}~\bibnamefont{Nowak}},
  \bibinfo{author}{\bibfnamefont{A.}~\bibnamefont{Misra}}, \bibnamefont{and}
  \bibinfo{author}{\bibfnamefont{K.~D.} \bibnamefont{Usadel}},
  \bibinfo{journal}{J. Magn. Magn. Mater.} \textbf{\bibinfo{volume}{240}},
  \bibinfo{pages}{243} (\bibinfo{year}{2002}{\natexlab{b}}).

\bibitem[{\citenamefont{Billoni et~al.}(2006)\citenamefont{Billoni, Tamarit,
  and Cannas}}]{Billoni06PB}
\bibinfo{author}{\bibfnamefont{O.~V.} \bibnamefont{Billoni}},
  \bibinfo{author}{\bibfnamefont{F.~A.} \bibnamefont{Tamarit}},
  \bibnamefont{and} \bibinfo{author}{\bibfnamefont{S.~A.}
  \bibnamefont{Cannas}}, \bibinfo{journal}{Physica B}
  \textbf{\bibinfo{volume}{384}}, \bibinfo{pages}{184 } (\bibinfo{year}{2006}).

\bibitem[{\citenamefont{Lederman et~al.}(2004)\citenamefont{Lederman,
  Ram\'{\i}rez, and Kiwi}}]{Lederman04PRB}
\bibinfo{author}{\bibfnamefont{D.}~\bibnamefont{Lederman}},
  \bibinfo{author}{\bibfnamefont{R.}~\bibnamefont{Ram\'{\i}rez}},
  \bibnamefont{and} \bibinfo{author}{\bibfnamefont{M.}~\bibnamefont{Kiwi}},
  \bibinfo{journal}{Phys. Rev. B} \textbf{\bibinfo{volume}{70}},
  \bibinfo{pages}{184422} (\bibinfo{year}{2004}).

\bibitem[{\citenamefont{Billoni et~al.}(2011)\citenamefont{Billoni, Cannas, and
  Tamarit}}]{Billoni11JPCM}
\bibinfo{author}{\bibfnamefont{O.~V.} \bibnamefont{Billoni}},
  \bibinfo{author}{\bibfnamefont{S.~A.} \bibnamefont{Cannas}},
  \bibnamefont{and} \bibinfo{author}{\bibfnamefont{F.~A.}
  \bibnamefont{Tamarit}}, \bibinfo{journal}{J. of Phys.: Condens. Matter}
  \textbf{\bibinfo{volume}{23}}, \bibinfo{pages}{386004}
  (\bibinfo{year}{2011}).

\bibitem[{\citenamefont{Kim and Stamps}(2001)}]{Kim01APL}
\bibinfo{author}{\bibfnamefont{J.-V.} \bibnamefont{Kim}} \bibnamefont{and}
  \bibinfo{author}{\bibfnamefont{R.~L.} \bibnamefont{Stamps}},
  \bibinfo{journal}{Appl. Phys. Lett.} \textbf{\bibinfo{volume}{79}},
  \bibinfo{pages}{2785} (\bibinfo{year}{2001}).

\bibitem[{\citenamefont{Chaikin and Lubensky}(1995)}]{Chaikin95}
\bibinfo{author}{\bibfnamefont{P.~M.} \bibnamefont{Chaikin}} \bibnamefont{and}
  \bibinfo{author}{\bibfnamefont{T.~C.} \bibnamefont{Lubensky}},
  \emph{\bibinfo{title}{Principles of Condensed Matter Physics}}
  (\bibinfo{publisher}{Cambridge University Press}, \bibinfo{address}{Cambrige,
  UK}, \bibinfo{year}{1995}).

\bibitem[{\citenamefont{Petracic et~al.}(2005)\citenamefont{Petracic, Li,
  Roshchin, Viret, Morales, Batlle, and Schuller}}]{Petracic05APL}
\bibinfo{author}{\bibfnamefont{O.}~\bibnamefont{Petracic}},
  \bibinfo{author}{\bibfnamefont{Z.-P.} \bibnamefont{Li}},
  \bibinfo{author}{\bibfnamefont{I.~V.} \bibnamefont{Roshchin}},
  \bibinfo{author}{\bibfnamefont{M.}~\bibnamefont{Viret}},
  \bibinfo{author}{\bibfnamefont{R.}~\bibnamefont{Morales}},
  \bibinfo{author}{\bibfnamefont{X.}~\bibnamefont{Batlle}}, \bibnamefont{and}
  \bibinfo{author}{\bibfnamefont{I.~K.} \bibnamefont{Schuller}},
  \bibinfo{journal}{Appl. Phys. Lett.} \textbf{\bibinfo{volume}{87}},
  \bibinfo{eid}{222509} (\bibinfo{year}{2005}).

\bibitem[{\citenamefont{Roshchin et~al.}(2005)\citenamefont{Roshchin, Petracic,
  Morales, Li, Batlle, and Schuller}}]{Roshchin05EPL}
\bibinfo{author}{\bibfnamefont{I.~V.} \bibnamefont{Roshchin}},
  \bibinfo{author}{\bibfnamefont{O.}~\bibnamefont{Petracic}},
  \bibinfo{author}{\bibfnamefont{R.}~\bibnamefont{Morales}},
  \bibinfo{author}{\bibfnamefont{Z.-P.} \bibnamefont{Li}},
  \bibinfo{author}{\bibfnamefont{X.}~\bibnamefont{Batlle}}, \bibnamefont{and}
  \bibinfo{author}{\bibfnamefont{I.~K.} \bibnamefont{Schuller}},
  \bibinfo{journal}{Europhys. Lett.} \textbf{\bibinfo{volume}{71}},
  \bibinfo{pages}{297} (\bibinfo{year}{2005}).

\bibitem[{\citenamefont{Radu and Zabel}(2008)}]{Radu08}
\bibinfo{author}{\bibfnamefont{F.}~\bibnamefont{Radu}} \bibnamefont{and}
  \bibinfo{author}{\bibfnamefont{H.}~\bibnamefont{Zabel}},
  \emph{\bibinfo{title}{Magnetic Heterostructures; Advances and Perspectives in
  Spinstructures and Spintransport; Series: Springer Tracts in Modern
  Physics}}, Vol. \bibinfo{volume}{227} (\bibinfo{publisher}{Springer-Verlag},
  \bibinfo{address}{Berlin Heidelberg}, \bibinfo{year}{2008}).

\end{thebibliography}

\end{document}